\documentclass[conference]{IEEEtran}
\usepackage{ifthen} 
\newboolean{showcomments}
\setboolean{showcomments}{true} 
\ifthenelse{\boolean{showcomments}}
   {

}

\IEEEoverridecommandlockouts

\usepackage{url}
\usepackage{tabularx}
\usepackage{booktabs}
\usepackage{cite}
\usepackage{amsmath,amssymb,amsfonts}
\usepackage{algorithmic}
\usepackage{enumitem}
\usepackage{graphicx}
\usepackage{textcomp}
\usepackage[table]{xcolor}
\usepackage[table]{xcolor}
\usepackage{array}
\usepackage{balance}
\usepackage{csvsimple} 
\usepackage{boldline} 
\usepackage{multirow} 
\usepackage{booktabs} 
\usepackage{tikz}
\usepackage{etoolbox}
\newcolumntype{L}[1]{>{\raggedright\let\newline\\\arraybackslash}p{#1}} 
\usepackage{roboto}

\definecolor{lightpurple}{RGB}{21, 11, 163}
\definecolor{lightgrey}{RGB}{211, 211, 211}

\tikzstyle{userbubble} = [rectangle, draw, fill=lightgrey!20, text width=.75\columnwidth, font=\roboto\scriptsize, minimum height=1cm, align=left, rounded corners]
\tikzstyle{assistantbubble} = [rectangle, draw, fill=lightpurple!20, text width=.75\columnwidth, font=\roboto\scriptsize, minimum height=1cm, align=left, rounded corners]
                           
\newenvironment{chatbubble}[2]
{
    \ifstrequal{#1}{User}{
        \begin{flushright}
            \begin{tikzpicture}
                \node[userbubble, anchor=east] at (0,0) {
                    \textbf{Human}: #2
                };
            \end{tikzpicture}
        \end{flushright}
  } 
  
  \ifstrequal{#1}{Assistant} {
    \begin{flushleft}
        \begin{tikzpicture}
            \node[assistantbubble, anchor=west] at (0,0){\textbf{Assistant}: #2};
        \end{tikzpicture}
    \end{flushleft}
  }

}
{}

\newcommand{\human}[1]{
    \chatbubble{User}{#1}
}
\newcommand{\assistant}[1]{
    \chatbubble{Assistant}{#1}
}
\newcommand{\brackets}[1]{
    \textless#1\textgreater
}
\def\BibTeX{{\rm B\kern-.05em{\sc i\kern-.025em b}\kern-.08em
    T\kern-.1667em\lower.7ex\hbox{E}\kern-.125emX}}

\begin{document}

\title{Prompts Matter: Insights and Strategies for Prompt Engineering in Automated Software Traceability
}

\author{
  \IEEEauthorblockN{Alberto D. Rodriguez}
  \IEEEauthorblockA{\textit{College of Engineering} \\
    \textit{University of Notre Dame}\\
    Notre Dame, Indiana \\
    arodri39@nd.edu}
  \and
  \IEEEauthorblockN{Katherine R. Dearstyne}
  \IEEEauthorblockA{\textit{College of Engineering} \\
    \textit{University of Notre Dame}\\
    Notre Dame, Indiana \\
    kdearsty@nd.edu}
  \and
  \IEEEauthorblockN{Jane Cleland-Huang}
  \IEEEauthorblockA{\textit{College of Engineering} \\
    \textit{University of Notre Dame}\\
    Notre Dame, Indiana \\
    JaneClelandHuang@nd.edu}
}
\maketitle

\begin{abstract}
Large Language Models (LLMs) have the potential to revolutionize automated traceability by overcoming the challenges faced by previous methods and introducing new possibilities. However, the optimal utilization of LLMs for automated traceability remains unclear. This paper explores the process of prompt engineering to extract link predictions from an LLM. We provide detailed insights into our approach for constructing effective prompts, offering our lessons learned. Additionally, we propose multiple strategies for leveraging LLMs to generate traceability links, improving upon previous zero-shot methods on the ranking of candidate links after prompt refinement. The primary objective of this paper is to inspire and assist future researchers and engineers by highlighting the process of constructing traceability prompts to effectively harness LLMs for advancing automatic traceability.
\end{abstract}


\section{Introduction}
\label{sec:intro}
The challenges of automating traceability have been well documented over the past two decades \cite{traceability_vsm,cm1,DBLP:books/daglib/p/LuciaMOP12,rath2018traceability}; however, achieving satisfactory degrees of accuracy across diverse datasets has been an ongoing challenge \cite{DBLP:conf/icse/Cleland-HuangGHMZ14,traceability_nn} that has inhibited its adoption in industry. The release of the Google's BERT model \cite{BERT} in 2018 introduced new possibilities for the field, transforming the once far off dream of automatic traceability into a reality for projects in common domains  \cite{bert_pl,bert_nl}. However, despite these improvements,  challenges such as highly-technical domain-specific terminology, low data availability for training, and lack of interpretability meant that automated tracing continued to under-perform in many projects and domains where trace links were still delivered at low degrees of accuracy \cite{MARO201885, DBLP:journals/computer/Cleland-HuangBCSR07}. In the present day, large language models (LLMs), such as GPT3 and Claude \cite{gpt3, claude}, offer the promise of further transformation in automated traceability, eliminating many of these problems and introducing new possibilities for the field. However, as of yet, there is no clear direction on how best to utilize LLMs for automated traceability. 

When we began the work for this paper, our initial aspiration was to discover the ``silver bullet'' prompt for automated traceability. Similar to previous approaches \cite{traceability_vsm, bert_nl, bert_pl}, the ``silver bullet'' would discern true candidate links from false ones across all projects and circumstances. While we identified a prompting approach that performed well across multiple projects, we concluded that the optimal prompting strategy depends on factors like available resources, the model being used, and the targeted usage scenario. Different LLMs exhibit distinct strengths and weaknesses and may require different prompts to achieve desired outcomes on the same data sets; compounding this, variance across versions of the same base model can alter performance on the same task \cite{chen_how_2023}. Moreover, top-performing models can be cost-prohibitive to many engineers and researchers. Despite LLMs’ capabilities, high variability persists across projects, prompts, and parameters. 

Therefore, by bringing attention to some of the obstacles we encountered while crafting out prompts, we hope to make researchers and practitioners aware of potential pitfalls when employing the models for traceability related tasks. Rather than merely showcase top results, we have chosen to elaborate on the process we followed to construct our prompts with the goal of inspiring other engineers who may wish to identify a prompt that best suits their needs.

In this paper, we seek to shed light on the following questions:
\begin{enumerate}
    \item Do LLMs possess knowledge necessary for tracing projects with technical domain-specific vocabularly?
    \item Can LLMs provide reasonable explanations for their decisions?
    \item If so, can these explanations be utilized to improve prompts?
    \item Can reasoning be used to improve responses?
    \item How can LLMs be leveraged to generate software traceability links?
\end{enumerate}

While much future work is needed in this area, we hope to aid future researchers and engineers by highlighting the process of constructing traceability prompts for leveraging LLMs effectively to advance automatic traceability. 

\begin{table*}[!t]
  \label{tab:datasets}
  \centering
  \caption{Datasets}
\begin{tabular}{p{2cm}p{5cm}p{5cm}cp{1.25cm}c p{1.25cm}cp{1.25cm}}

    \toprule
    \textbf{Project Name} & \textbf{Description} & \textbf{Artifacts} & \textbf{Children} & \textbf{Candidates} & \textbf{True} \\
    \midrule \midrule
    \multirow{2}{*}{\textbf{CM1 \cite{cm1}}} & The requirements for an instrument a part of NASA's Metric Data Program (MDP). & High-Level Requirements $\rightarrow$ Low-Level Requirements & \multirow{2}{*}{53} & \multirow{2}{*}{265} & \multirow{2}{*}{13} \\ \midrule
    \multirow{3}{*}{\textbf{iTrust \cite{Meneely2011_itrust}}} & Open-source electronic health record system. Created at North Carolina State University as a part of a software engineering course. & \multirow{3}{*}{Requirements $\rightarrow$ Java Classes} & \multirow{3}{*}{227} & \multirow{3}{*}{1135} & \multirow{3}{*}{13} \\ \midrule
    \multirow{3}{*}{\textbf{Dronology \cite{clelang_huang_dronology}}} & \multirow{3}{6cm}{A system for managing the navigation of \\ UAVs and their communication to the ground control station.} & \textbf{NL}: Requirements $\rightarrow$ Design Definitions & 99 & 495 & 4 \\
    & & & & & \\ 
    & & \textbf{PL}: Design Definitions $\rightarrow$ Java Classes & 458 & 2290 & 48 \\ \bottomrule
  \end{tabular}
  {\newline \newline \raggedright Describes the artifact types in each dataset, the number of children per query, the resulting candidate links across all queries, and how many of those candidates were true links. Dronology is split into two datasets, DronologyNL and DronologyPL, to focus on traces between natural language artifacts and between natural language and programming language artifacts respectively.}
   \vspace{-12pt}
\end{table*}

\section{Related Work}
\label{sec:related}
Effective automated software traceability has many benefits for software engineering, and several approaches have therefore been proposed to address its challenges. In recent years, the emergence of LLMs, such as GPT-3 and Claude, has shown promise for automating software traceability and mitigating the limitations of previous methods. In this section, we discuss the relevant works that have explored the use of large language models and the subjectivity of trace establishment in the context of software traceability.

Early work in automated traceability relied on classical natural language processing (NLP) techniques such as the vector space model (VSM) and latent semantic indexing (LSI) to establish traceability links between software artifacts based on their textual similarity \cite{traceability_vsm, traceability_lsi}. In the 2010s, deep learning techniques such as long short-term memory networks (LSTMs) and gated recurrent units (GRUs) were applied to improve traceability performance. Researchers used these neural networks to learn distributed representations of software artifacts and match them based on semantic similarity \cite{traceability_nn}. Around 2018, pretrained language models and transformers revolutionized the field. Models like Google's BERT allowed researchers to generate contextualized embeddings of software artifacts and achieve state-of-the-art results in automated traceability tasks \cite{bert_pl, bert_nl}. Transformer language models then grew exponentially larger and more powerful, culminating in GPT-3 and models with hundreds of billions of parameters. GPT-3 demonstrated human-level language understanding with 175 billion parameters, achieving startling fluency and few-shot learning capabilities\cite{gpt2, gpt3, palm}. GPT-4 continues to push the limits of LLMs, scoring in the top 10\% of the BAR exam \cite{gpt4}.

In the domain of software engineering, efforts have been made to leverage large language models for various software engineering tasks including code generation, summarization, and enhancement \cite{DBLP:journals/corr/abs-2107-03374, Sridhara_G_Mazumdar_2023}. Although prompt-engineering is a relatively new area of exploration, some prior work has been done on how best to instruct models for various tasks. Researchers have identified different prompt patterns and techniques that tend to produce the best results - many of which are employed in this paper \cite{white_prompt_2023, ekin_prompt_2023}. Additionally, prompt engineers have crafted prompts for a variety of tasks, including classification \cite{mayer_prompt_2023, han_ptr_2021} and ranking \cite{qin_large_2023}, both of which we utilize in this paper.

However, there has not been extensive evaluation of the potential for large language models in automated software traceability. To address this gap, we conducted a preliminary investigation using Claude, an LLM developed by Anthropic, to predict trace links between software artifacts. We outlined our two approaches for trace link prediction: classification and ranking. The evaluation of our approaches will be discussed in the following section.

\section{Experimental Setup}
\label{sec:design}
For the preliminary investigation reported in this paper, we analyzed three software engineering datasets: CM1, iTrust, and Dronology. We selected these datasets to span natural language and programming language artifacts as well as diverse application domains (embedded systems, healthcare, UAVs). 

For each dataset, we selected only a subset of its data to use in our study in order to increase the depth of our analysis, reduce run-time, and decrease cost. To select the links, we first calculated the number of children artifacts traced to each parent and then identified the minimum, maximum and median number of links. Using these categories, we identified five parent artifacts: one with the fewest child links, three with the median number of child links, and one with the maximum number of child links. In cases where multiple parent artifacts tied for the minimum, median, or maximum, we randomly sampled from those tied parents. This allowed us to create a set of trace queries that were representative of the project's link distribution of its trace queries. Table \ref{tab:datasets} describes the selected queries for each system noting the parent and child types, the number of potential trace links (candidates), and the number of those links that were actually true.

Prior to the start of our experiments, we tested OpenAI's \emph{text-davinci-003} model for predicting trace links, and found that, while it required slightly different prompts, it had comparable capabilities to Anthropic's Claude instant model (\emph{claude-instant-v1}). Due to its lower cost and increased speed, we selected Claude for the remainder of our experiments. We also explored utilizing embeddings to compute similarity scores between artifacts, similar to the original Vector Space Model (VSM) approaches \cite{traceability_vsm}. We examined the ada-embedding model developed by OpenAI (\emph{text-embedding-ada-002}), however, the results obtained from this investigation did not show a significant advantage over VSM. Therefore, we decided to leverage the generative capabilities of the models for trace link predictions within this paper. Nevertheless, we acknowledge the need for future endeavors to conduct a more comprehensive analysis of the advantages and disadvantages associated with utilizing embeddings for generating trace links.

 Additionally, we obtained summaries of all code artifacts to use in our experiments. We accomplished this by prompting the model to provide a several sentences focusing on the high-level functionality of the code. Although this removed some information, the resulting summaries contained most of the relevant details and reduced the number of tokens required for each tracing prompt.

For our first approach, we prompted the model to classify every source and target artifact pair. Each prompt followed a similar format, consisting primarily of a question and instructions to answer `yes' or `no', followed by the content of the source artifact numbered as `1' and the target artifact numbered as `2'. When a prompt directly referenced the source or target in the question, it used (1) to indicate the source or (2) to indicate the target, corresponding to the numbers of the artifact content (e.g., ``Is (1) related to (2)?"). Each question was posed such that an answer of `yes' was indicative of a link between the answers, while `no' indicated that the artifacts were not linked. The resulting candidate links are then evaluated against the ground truth links using common classification metrics such as precision and recall. 

Precision is the ratio of the number of correctly identified relevant trace links to the total number of trace links identified by the system. Recall, on the other hand, measures the ratio of the correctly identified relevant trace links to the total number of relevant trace links in the system. This is shown below where TP is the true positives, FP is false positives, and FN is false negatives.
\begin{align*}
\text{Precision} &= \frac{\text{TP}}{\text{TP} + \text{FP}} &
\text{Recall}    &= \frac{\text{TP}}{\text{TP} + \text{FN}}
\end{align*}

For our ranking approach, we prompted the model to rank all targets for each source artifact. In this case, the model was given the content of the source artifact and the ID and body of each target, separated by newlines. The model was instructed to return the artifact IDs in the order of relevance (from most to least) in a comma delimited list. Given the non-deterministic nature of responses from each model, there were times when the model neglected to include some artifact IDs. This problem was unique to the ranking task, as the model correctly output yes and no each time for the classification task. In these cases we randomly appended the missing ids to the end of the list for our evaluation. We calculate the Mean Average Precision of these rankings to showcase their performance. It provides a measure of the effectiveness of the ranking algorithm in identifying relevant trace links between software artifacts. To calculate MAP, the precision is computed at different levels of recall. The average precision is then calculated as the average of the precision values at each recall level. Finally, the mean of the average precision values across trace queries is taken to obtain the MAP score. The equation for MAP is obtained by taking the mean of the average precision values across different queries or datasets:

\begin{equation*}
\text{MAP} = \frac{1}{N} \sum_{i=1}^{N} \text{Average Precision}_i
\end{equation*}

where $N$ is the number of queries or datasets.

Throughout our process of generating trace-links, we have several conversations with the model to test its prior knowledge, understand its responses, and to brainstorm potential prompts and improvements to prompts. We include many of these in our paper. It is important to note that these exchanges occurred independently - the model could not reference previous conversations when responding to subsequent questions or when making new traceability classifications. Each conversation represented an isolated context, rather than a continuing thread. This methodology enabled us to systematically probe the model's knowledge without reliance on prior memorization.




\section{Analysis and Prompt Discovery}
\label{sec:results}
Due to space constraints, we focus our analysis on our prompt engineering process for the CM1 dataset. While we followed a similar prompting process with other datasets, we found that many of the challenges we faced were exemplified in CM1. We specifically chose CM1 to focus on because it has posed a unique challenge to researchers in the past \cite{bert_pl, bert_nl}. In the end, we show the performance of our best approach on all the datasets to highlight that many of the lessons we learned from CM1 were transferable to the other datasets as well.

\subsection{Domain-specific Knowledge}
Before running trace-link prediction with the models, we first wanted to understand what prior knowledge the model had about the system being evaluated. Previously, BERT models often required extensive pre-training on domain-specific documentation to interpret the vocabulary within a project \cite{bert_pl, bert_nl}. CM1, in particular, includes a myriad of jargon and acronyms that have long presented a challenge to previous methods of automatic traceability \cite{cm1, rodriguez2023understanding}. Since one strength of LLMs is their ability to perform well without any training \cite{gpt3, gpt4}, we wanted to explore whether Claude possessed enough prior knowledge to succeed on the CM1 dataset. To do this, we had the following conversation with it about one of the CM1 artifacts containing a large number of acronyms.

\human{What does this artifact mean? \\
The DPU-TMALI shall utilize SCM-DCI-SR, along with ERRNO provided by DPU-DCI to decode errors and place them on an error queue for DPU-CCM.}

\assistant{
... Some guesses at the meaning:  \newline \newline DPU may stand for Data Processing Unit ... SCM-DCI-SR seems to be some kind of protocol or interface used for communication between DPU-TMALI and DPU-DCI. \newline \newline ERRNO is a standard way for software components to return error codes. So DPU-DCI seems to be providing error codes to DPU-TMALI ... DPU-TMALI is receiving error data from DPU-DCI, decoding or interpreting those errors, and then queuing them up to be sent to DPU-CCM. \newline \newline DPU-CCM seems to be a component that is receiving error data, perhaps for monitoring purposes...}

As shown in the conversation above, the model was able to leverage its general knowledge to reason about what each component did from only the artifact content. Although we did not extensively investigate the model's knowledge of each acronym or vocabulary word in the dataset, this example shows that Claude can interpret the meaning of complex acronyms in some cases where previous models could not without fine-tuning \cite{bert_nl, bert_pl}. 

\subsection{Classification Prompt Discovery}

Next, we began the process of creating a prompt to use for our trace link prediction. As discussed previously, we provide highlights only for our experiments with CM1 but performed a similar analysis across datasets. Results for each prompt can be found in Table \ref{tab:cm1_prompt_metrics}.
Our initial prompt for the model was general:
\human{Is (1) related to (2)? Answer ‘yes’ or ‘no’. \\}

The content of the two artifacts was included in the prompt following the format explained in Section \ref{sec:design}. On CM1, this prompt resulted in the model correctly identifying 6 out of 13 links and mispredicting 27 of the 225 true negative links as positive links for precision of 18\% and recall of 46\%.

In order to understand why the prompt failed on over half of the true links, we randomly selected one of the false negatives and prompted the model to explain why it responded `no' for the link. For reference, the artifacts in the link are shown in Table \ref{tab:investigated_links} in the row for Link \#1.  

\assistant{...The first paragraph is very high level, while the second paragraph goes into more technical implementation details. If they were related, you would expect a more consistent level of technical depth...}

Inspired by this response, we inserted ``Ignoring the different abstraction levels of the artifacts..." before our question. Interestingly, this link was correctly predicted when we used the modified prompt. Throughout our experiments, this would be a re-occurring problem where the model often failed to predict links between artifacts with different levels of details/abstractions. We therefore began instructing it to ignore abstraction levels in many of our following prompts. When we re-ran this modified prompt on the dataset, it correctly identified 7 out of the 13 links (one more than before) but also predicted 86 false positives. We again prompted it to explain its decision for one of the remaining false negatives (Link \#2 in \ref{tab:investigated_links}). Its response and our subsequent questioning are shown below:

\assistant{...while there is a small overlap in the error handling mention, the main substance and purposes of the artifacts are distinct. I would still assess them as not strongly related...}

\human{Would there be a traceability link between artifact 1 and 2?}

\assistant{There could potentially be a weak traceability link between the two artifacts based on the minor overlap in error handling/queueing...}

We found that asking it directly if the artifacts were linked received a better response, and further prompting revealed that Claude had a relatively robust understanding of traceability, thus we decided to update our prompt as follows: 
\human{Is there a traceability link between (1) and (2)?}

 We suspected that this prompt would be more specific, potentially eliminating some of the false positives, and this was indeed the case with Claude only predicting one false positive. However, it now correctly identified only 4 of the 13 links. This time when questioned about why it incorrectly believed Link \#3 (Table \ref{tab:investigated_links}) was un-traced, Claude's confusion seemed to come from it not realizing the artifacts were talking about the same system:

\assistant{
 ...The first requirement mentions specific tasks...that need to report heartbeats, while the second requirement talks about tasks in general without specifying which tasks. So it's not clear if the tasks mentioned in the first requirement are the same as those referred to in the second...}

We found this to be another common point of confusion for Claude so we updated our prompt to include ``Below are artifacts from the same software system" at the start of our prompt. 
\human{Below are artifacts from the same software system. Is there a traceabilty link between (1) and (2)?}
With this additional information, Claude identified all of the project trace-links but wrongly predicted 98 negative links to be positive. Since Claude now mistakenly believed many artifacts were linked, we randomly selected a false positive (Link \#4 in Table \ref{tab:investigated_links}) to investigate further. This lead us to gain the following insight into the misprediction: 
 
\assistant{
Yes, there is a traceability link between (1) and (2). (1) mentions that the DPU-TMALI places errors on an error queue for the DPU-CCM. \newline \newline (2) describes the error collection and reporting mechanism of the DPU-CCM, specifically referring to enqueueing errors. \newline \newline So (2) provides an implementation detail for the error queueing functionality briefly mentioned in (1). Therefore, there is a traceability link between these two artifacts in the software system.}

We found nothing in Claude's response to be inherently wrong, so we speculated that the mistake came not from a misunderstanding but rather from a misalignment between Claude and the tracers of CM1 on what should be considered a trace-link. Thus, we began to generate prompts that focused on the specific relationships between the requirements and regulations of CM1.
\human{Below are artifacts from the same software system. Ignoring different levels of abstraction, can (2) be decomposed from (1)? Answer `yes' or `no'.}

This led to the model identifying 8 of the trace links with only 14 false positives. These missed trace links seemed to result from a confusion surrounding the model's understanding of decomposition, where the model believed that no new information could be added to a decomposed requirement. This was indicated by its response below for Link \#1:

\assistant{... So (2) contains new information that cannot be derived or decomposed from (1) alone. The answer is 'no'. }

Replacing the question with ``does (2) fulfill (1)?" resulted in the model missing 7 links for a similar reason, including Link \#1 again:

\assistant{...(2) lacks critical details and scope needed to conclusively fulfill (1)...}

Despite missing 7 links, the model was able to correctly rule out all but 3 true negatives. Although both of these prompts were too restrictive (i.e. not all the true links were captured), they seemed to align Claude with at least some of the relationships established in CM1 as indicated by its success at eliminating false links.

Some other prompts we tried included:
\begin{enumerate}[label=\arabic*.]
    \item \textit{are (1) and (2) part of the same feature and share functionality?}
    \item \textit{does (2) implement system functionality from (1)?}
    \item \textit{are (1) and (2) dependent on the same system capability?}
\end{enumerate}
where each prompt began with ``Below are artifacts from the same software system. Ignoring different levels of abstraction" and concluded with "Answer `yes' or `no'." Each subsequent prompt successfully predicted one more positive link than the one before, but led to an increasing number of false positives ranging from 28 all the way to 48 for \#3. 

Our final prompt of this structure was a combination of several previous prompts that had shown promise:

\human{Below are artifacts from the same software system. Ignoring different levels of abstraction, can (2) be one of the hierarchical decompositions of (1)? Consider whether (2) implements a part of (1).}

By adding ``one of the hierarchical" and ``a part of", we hoped to make the prompt less restrictive so that the model would allow new information to be added to decomposed artifacts. This time the model correctly predicted all but one true link. Its 44 false positives still fell short of our goal of less than 20 but were significantly less than the 98 returned in the previous approach that achieved these high levels of recall.

\subsection{Introduction of Chain-of-Thought Reasoning}
As our prompt evolved, we saw improvements in the model's link-prediction abilities, but the results were still falling short of our goals. It seemed evident that we needed to approach the task differently. As a result, we decided to break down the task into steps to aid the model in reasoning about the relationship between the artifacts before arriving at a final decision. This process, known as "chain of thought", has previously shown to be effective in helping LLMs solve complex tasks \cite{wei_chain--thought_2023}. To test the impact of incorporating reasoning into the process, we reran our original prompt with this additional step at the beginning of the prompt:

\human{You are working on a software system and are given a high-level requirement and a low-level requirement. Your job is to determine if they are traced. Give one reason why (1) might be related to (2) enclosed in \brackets{related}\brackets{/related} and one reason why (1) might be un-related to (2) enclosed in \brackets{unrelated}\brackets{/unrelated} but ignore levels of abstraction. Then based on these reasons, do you think (1) is related to (2)?  Answer should be 'yes' or 'no' enclosed in \brackets{answer}\brackets{/answer}.}

Having it enclose each separate response in xml tags allowed us to easily parse its answer despite the added complexity. An example of the model's response for link \#5 in Table \ref{tab:investigated_links} is shown below:

\assistant{\brackets{related} Both specify requirements related to processing commands within a certain time \brackets{/related}
\brackets{unrelated} Requirement 1 specifies a time constraint while requirement 2 provides a general description of a command and control subsystem\brackets{/unrelated} \brackets{answer}yes\brackets{/answer}} 

Recall increased by nearly 50\% while precision rose by 14\%. With the added reasoning step, the model missed only 1 true link and misidentified just 25 false positives. With both questions (1) and (3) from Table \ref{tab:cm1_prompt_metrics}, the introduction of incremental reasoning helped the model find a higher portion of the true connections in the data. Due to time constraints, we were unable to test the reasoning on the remaining questions but we believe this is an interesting avenue for future work.

Encouraged by this initial success, we decided to have the model answer each of our questions as intermediate steps before finally determining whether the artifacts were related. We hoped this approach would help the model explore different ways in which the artifacts could be connected. It also allowed us to use a simple ranking system in which more `yes' responses would increase the likelihood that the artifacts were linked. By quantifying the model's degree of support for a relationship through the ranking system, we could evaluate not just whether it predicted a link but also how confident it was in that prediction based on the reasoning exhibited in its responses. 

\human{I am giving you two software artifacts from a system.
Your job is to determine if there is a traceability link.
Answer whether (2) implements a part of (1) with yes or no enclosed in \brackets{implements}\brackets{/implements}.
Answer whether (2) is a hierarchical decomposition of (1) with yes or no enclosed in \brackets{decomposed}\brackets{/decomposed}.
Answer whether (2) fulfills (1) with yes or no enclosed in \brackets{fulfills}\brackets{/fulfills}.
Answer whether (2) and (1) are part of the same feature and shares functionality with yes or no enclosed in \brackets{feature}\brackets{/feature}.
Answer whether (2) and (1) are dependent on the same system capability with yes or no enclosed in \brackets{capability}\brackets{/capability}.
Use your answers to give one reason why (1) might be related to (2) enclosed in \brackets{related}\brackets{/related}
and one reason why (1) might be un-related to (2) enclosed in \brackets{unrelated}\brackets{/unrelated}
Now answer is (1) related to (2) with yes or no enclosed in \brackets{traced}\brackets{/traced}.}

\subsection{Ranking Prompt Discovery}
Despite not outperforming other classification prompts, ranking the artifacts by the number of `yes' and `no' answers, did provide the opportunity to establish a threshold retrospectively, allowing us to categorize items based on the strength of the model's prediction instead of relying on a single yes/no choice. This, combined with Claude's new 100k context window, inspired us to experiment with an entirely new strategy.

For our next experiment, we gave Claude the following instructions:
\human{
\# Task \newline 
Rank all related artifacts from most to least related to the source.\newline \newline
Source: [SOURCE ARTIFACT] \newline \newline \# Artifacts \newline \newline \brackets{artifact}\newline 
\brackets{id}...\brackets{/id}\newline
\brackets{body}...\brackets{/body}\newline
\brackets{/artifact} \newline \newline
\# Instructions \newline Rank the artifact bodies from most to least relevant to the source. Provide the ranked artifacts as comma delimited list of artifact ids where the first element relates to the source the most and the last element does so the least. 
}

\begin{table*}
\centering
\caption{Classification Metrics for CM1  Prompts}
\label{tab:cm1_prompt_metrics}
\begin{tabular}{cp{11cm}|cccccc}
\toprule
ID & Prompt & Precision & Recall & TP & TN & FP & FN \\
\midrule\midrule
1 & \textbf{Is (1) related to (2)?} & 18\% & 46\% & 6 & 225 & 27 & 7 \\  
& & \cellcolor{gray!20}32.4\% & \cellcolor{gray!20}92.3\% & \cellcolor{gray!20}12 & \cellcolor{gray!20}227 & \cellcolor{gray!20}25 & \cellcolor{gray!20}1 \\
\midrule

2 & \textbf{Ignoring the different abstraction levels of the artifacts, is (1) related to (2)?} & 17\% & 54\% & 7 & 218 & 34 & 6 \\
\midrule

3 & \textbf{Is there a traceability link between (1) and (2)?} & 80\% & 31\% & 4 & 251 & 1 & 9 \\ & & \cellcolor{gray!20}40\% & \cellcolor{gray!20}46.2\% & \cellcolor{gray!20}6 & \cellcolor{gray!20}243 & \cellcolor{gray!20}9 & \cellcolor{gray!20}7 \\
\midrule

4 & \textbf{Below are artifacts from the same software system, is there a traceability link between (1) and (2)?} & 12\% & 100\% & 13 & 154 & 98 & 0 \\
\midrule

5 & \textbf{Below are artifacts from the same software system. Ignoring different levels of abstraction, can (2) be decomposed from (1)?} & 36\% & 62\% & 8 & 238 & 14 & 5 \\
\midrule

6 & \textbf{Below are artifacts from the same software system. Ignoring different levels of abstraction, does (2) fulfill (1)?} & 67\% & 46\% & 6 & 249 & 3 & 7 \\
\midrule

7 & \textbf{Below are artifacts from the same software system. Ignoring different levels of abstraction, are (1) and (2) part of the same feature and share functionality?} & 32\% & 54\% & 7 & 237 & 15 & 6 \\
\midrule

8 & \textbf{Below are artifacts from the same software system. Ignoring different levels of abstraction, does (2) implement system functionality from (1)?} & 22\% & 77\% & 10 & 216 & 36 & 3 \\
\midrule

9 & \textbf{Below are artifacts from the same software system. Ignoring different levels of abstraction, are (1) and (2) dependent on the same system capability?} & 19\% & 85\% & 11 & 204 & 48 & 2 \\
\midrule

10 & \textbf{Below are artifact from the same software system. Ignoring different levels of abstraction, can (2) be one of the hierarchical decompositions of (1)? Consider whether (2) implements a part of (1).} & 22\% & 92\% & 12 & 208 & 44 & 1 \\
\midrule

11 & \textbf{Combining all questions and chain-of-thought reasoning.} & \cellcolor{gray!20}37.9\% & \cellcolor{gray!20}84.6\% & \cellcolor{gray!20}11 & \cellcolor{gray!20}234 & \cellcolor{gray!20}18 & \cellcolor{gray!20}2 \\
\bottomrule
\end{tabular}
{\newline \newline \raggedright Rows in gray use chain-of-thought to make their final trace classifications.}
\vspace{-12pt}
\end{table*}

By providing the model with more context about the system in the prompt and allowing it to compare all targets when making its decision, we hoped to see a performance boost. Unfortunately, the task was not as simple as we had hoped, and we, like previous researchers, identified another nuance with the prompts - order matters \cite{qin_large_2023}. When we presented the target artifacts in a random order, performance was barely above random; however, ordering artifacts that were more likely to be linked at the top, delivered significantly higher performance. It seemed that unless there was some pattern already established, the task would overwhelm the model. Because of this, we decided to rank the target artifacts based on their VSM similarity to the source. Then, we presented the model with targets in this order. With this initialization, the model improved upon the original VSM ranking.  Furthermore, While discussions throughout the paper have focused on the CM1 dataset, we applied this approach to the three other datasets presented in Table \ref{tab:datasets} and report results for all four datasets in Table \ref{tab:performance_metrics}.

\subsection{Summary of Results}
Overall, our results demonstrated that the ranking task could be a useful approach to automated traceability, but it may require additional steps and further prompt refinement to reach the necessary performance. In the future, we plan to explore ways of decomposing the overall task into simpler, incremental steps to reduce complexity for the model as we did for the classification task. It should also be noted that the ranking task necessitated a large context window, which may pose a challenge for certain open-source models. Consequently, classification remains a valuable alternative when ranking is infeasible. Furthermore, classification opens up avenues for diverse applications of traceability, such as ``trace views" that we discuss further in Section \ref{sec:conclusion}.

\section{Threats to Validity}
\label{sec:threats}
While this initial study provides promising evidence that prompt engineering can enhance LLMs for software traceability tasks, several threats could limit the validity of our findings. First, we evaluated only three open-source projects and only provide a detailed analysis of one, limiting the generalization of our findings. However, we selected projects that spanned multiple domains, artifact types, and sizes to improve generalizability. We also constructed trace queries that were representative of their parent distribution. Second, existing traceability datasets are typically incomplete, as truly considering every candidate link in a project grows $\mathcal{O}(n^2)$ with the number of artifacts. The LLMs identified potential missing traces, but we could not fully validate their accuracy without a project expert. Third, our study used a limited set of LLMs which may not represent the full space of the current state-of-the-art. However, we chose the leading LLMs from our initial explorations with publicly available commercial models. Clearly, there are many extension to this study considering more datasets, different LLMs, and other prompt engineering methods. We leave the full exploration of the problem space to future work and focus on showing the potential these models have towards advancing automated software traceability.

\section{Conclusions and Future Directions}
\label{sec:conclusion}
\begin{table}[b]
    \centering
    \caption{Mean Average Precision of  Ranked Target Artifacts }
    \begin{tabular}{lcc}
        \toprule
        \textbf{Dataset} & \textbf{Ranking by VSM} & \textbf{Ranking by VSM + LLM} \\
        \midrule
        CM1 & 70.7\% & 79.4\% \\
        iTrust & 44.5\% & 44.2\% \\
        Dronology (NL) & 82.9\% & 100\% \\
        Dronology (PL) & 23.2\% & 30.8\% \\
        \bottomrule
    \end{tabular}
    \label{tab:performance_metrics}
\end{table}

Throughout our experiments, we addressed multiple questions and derived several key takeaways regarding using LLMs for trace-link prediction.

\subsection{Key Takeaways}

\begin{itemize}
\item Small modifications to prompts can lead to significant differences in model outputs, emphasizing the importance of carefully crafting prompts.

\item The performance of a given prompt in comparison to alternative phrasings can vary across datasets and models, though some general techniques like chain-of-thought reasoning tend to produce a more consistent performance.

\item LLMs frequently identify different artifact relationships by than those selected by human tracers. Prompts should specify the targeted usage of the traceability links (e.g. change impact analysis, hierarchical composition) to better align the model's output with the desired outcome.

\item Specifying the targeted usage has the additional benefit of opening the door for creating different trace views - a possible advantage over purely similarity-based methods.


\item Requiring models to show intermediate reasoning steps boosts performance on some tasks and builds in explanations into the decision making process. This is useful to both to those establishing the trace links and those using them. 

\item List ranking style prompts are highly sensitive to the order of artifacts presented in the prompt. This variability was mitigated by pre-sorting by VSM scores.

\item Overall, carefully tailored prompts are needed to harness the versatility of LLMs for the task of traceability and to produce outputs that are consistent with the goals of traceability engineers and researchers.

\end{itemize}

Throughout this process, one of our biggest takeaways was how minor adjustments to prompts could have dramatic impacts on the results. Subtle changes, such as pluralizing words, interchanging prepositions, or reordering phrases, could alter the outcomes. These findings underscore the inherent challenge of engineering robust prompts. In future research, we aim to explore strategies that mitigate such variability and delve into the effectiveness of different prompts across different models.

Further, due to the limited number of trace queries we analyzed per dataset as well as our integration of chain-of-thought, we were able to review trace predictions in depth. Interestingly, we were often surprised by the strength of many false positives, forcing us to re-think the accurate and complete nature of these datasets. Reviewing predictions for even our smallest subset (265 combinations) became an arduous task.  In reality, industrial projects range from 50K to 500K potential trace links, making it extremely challenging to have complete and standardized tracing practices. However, examining the predictions of a few selected trace links may still provide traceability experts with the insights they need to refine prompts in a way that improves performance across the project.

\subsection{Do LLMs possess knowledge necessary for tracing projects with domain-specific vocabulary?}
Our conversations with Claude revealed that it contained sufficient knowledge to draw many correct conclusions about the CM1 system, irrespective of the acronyms or jargon used. Furthermore, we were able to obtain high MAP scores without performing any additional pre-training.
Nevertheless, we plan to experiment with pre-training in the future to see if it can provide a performance boost. Additionally, we hope to test the model's knowledge on a wider range of datasets. It is important to note that since the datasets in this paper were all publicly available at the time of the model's creation, we cannot eliminate the possibility that the model had previous exposure to them. Thus, we are particularly interested to see how the model performs on an entirely new dataset.

\subsection{Can LLMs provide reasonable explanations for their decisions?}
By probing the model to elicit explanations for many of its mispredictions, we found that it could provide an in-depth analysis of its decision. Whether or not these explanations are accurate reflections of the reasoning behind the model's decision is beyond the scope of this paper, but we did find that when we adjusted the prompts based on the model's explanation, we were often able to change its answer.

\subsection{If so, can these explanations be utilized to improve prompts?}
The ability to alter the model's decision by using its explanations proved to be a useful tool for improving prompts. Engaging in conversations with the model enabled an increased understanding of its interpretation of a given prompt, facilitating an iterative approach to refine prompts. Gradually adjusting the prompts in this way can be used to find a prompt that better aligns the model's understanding with the objectives of the tracer.

\subsection{Can reasoning be used to improve responses?}
By asking the model to formally articulate its thinking in response to probing questions, the model was able to make a more well-informed final judgment about the relationship between the artifacts in the classification. This also offers the advantage of allowing the task to be broken down into smaller pieces, where the model first evaluates the relationship between the artifacts and then makes a final decision. Further, chain-of-thought reasoning has the potential to improve the ranking task and should be evaluated in future work.

\subsection{How can LLMs be leveraged to generate
software traceability links?}
In our experiments, we explored two different tasks which could be used to predict trace links from pairs of software artifacts: classification and 
ranking. While ranking allows for a nuanced expression of confidence in a prediction, classification  offers the advantage of needing a smaller context window and enables the discovery of diverse relationship types. By adapting our prompts to describe various relationships, we captured distinct links. For instance, when inquiring whether two artifacts were part of the same feature, we discovered different links than when asking if they shared functionality. This can be used to present multiple ``views" of traceability, where each view highlights different relationships within the system. This may be particularly valuable for change propagation where the prompt can focus on determining whether a modification to one artifact necessitates a change in the other. Additionally, multiple prompts may be combined to capture the many different relationships present in the project. This presents an avenue for future investigation.

An alternative way in which LLMs can be used for trace link prediction is by comparing the similarity of artifact embeddings. As mentioned previously, we opted not to explore this method in this paper, but future works might benefit from comparing this approach to those discussed in this paper.

\subsection{Concluding Remarks}
Overall, our experiments demonstrated that large language models show promise for tracing software systems. As opposed to previous approaches for automated traceability, LLMs can perform well without pre-training and are able to offer detailed explanations of their decisions. These explanations are not only useful for helping an engineer make an informed decision about a trace-link but can guide the process of selecting an appropriate prompt for the tracing task. Through iterative prompt refinement, the models can be used to classify trace links and establish a diverse set of relationships between project artifacts. The models are also capable of ranking target artifacts based on how related they are to a source artifact, albeit with aid from VSM. Ranking can allow engineers to sift through a prioritized list of candidate links and potentially reducing the review time required.

While this paper showcases the power of LLMs for traceability, it also highlights many of the lingering challenges in engineering effective prompts for the models. Careful tailoring of prompts can help to reach high performance for each project but this was ultimately a time-consuming task that may not always be feasible. Although the community might one day discover a "silver bullet" prompt, a more practical path forward may be to identify common patterns that make prompts most effective for certain projects and tracing objectives. Discovering such patterns could enable partially automating this process so that it can be seamlessly integrated into current traceability workflows. There remains much future work that must be done to gain a comprehensive understanding of how LLMs can best be utilized to enhance the field of traceability.

\section*{Acknowledgement}The work in this paper has been partially funded by USA National Science Foundation Grants \# SHF-1901059, SHF-1909007, and PFI-TT-2122689.

\begin{table*}
  \centering
  \caption{Investigated Links}
  \small
  \begin{tabularx}{\textwidth}{c L{4cm} X}
    \toprule

    \multicolumn{1}{c}{ID} & \multicolumn{1}{c}{Source} & \multicolumn{1}{c}{Target}\\
    \midrule \midrule
    1 & The DPU-CCM shall implement a mechanism whereby large memory loads and dumps can be accomplished incrementally. & Memory Upload and Download Handling Data can be uploaded to several types of locations, including:
        \begin{itemize}
          \item DRAM
          \item EEPROM
          \item Hardware registers
          \item EEPROM filesystem
        \end{itemize}
        The D-MEM-DAT-UPLD command specifies the target location. If the destination is the EEPROM filesystem, a "block number" is provided in lieu of a memory address, which is used by the DPU FSW to formulate a filename of the form \textit{eefs1:DPU\_blk.\#\#}, where \#\# is the block number. In this case, once the entirety of the uploaded data is received by the DPU FSW, the uploaded data is then written to that file in the EEPROM filesystem. If a file already exists with that name, it is overwritten. The EEPROM filesystem can be reinitialized using the command D-MEM-DISK-INIT. \\
    \midrule
    2 & The DPU-TMALI shall utilize SCM-DCI-SR, along with ERRNO provided by DPU-DCI to decode errors and place them on an error queue for DPU-CCM. & Control and Monitoring the CCM Control Task initializes the DPU FSW. It is the responsibility of the CCM Control Task to establish a successful boot. It does so by blocking on temporary semaphores, each with a 5 second timeout, after spawning the SCU Interface Task and the CCM Command Task. If both of these tasks report a successful initialization by giving the semaphore, the CCM Control Task toggles the BC\_INDEX parameter in EEPROM to indicate a successful boot. If either task does not report a successful initialization, the CCM Control Task disables the watchdog strobe to effect a reboot of the DPU. The rationale for selecting the successful initialization of these two tasks as the definition of a successful boot is that the DPU FSW requires these tasks, as a minimum, to establish ground contact and provide commandability. Once this initialization is complete, the task blocks on a binary semaphore which is given by the SCUI Command ISR upon arrival of the 1 Hz Clock Message. In the event a Clock Message does not arrive, the semaphore will time out after 1.5 seconds. The CCM Control Task remains alive to create and transmit DPU housekeeping at the appropriate intervals, perform various periodic processing tasks, and to process memory dump commands. The final call to ccmErrEnq() is performed in order that if an error occurs in an interrupt service routine, a global variable is set to the value of the errno which is then enqueued into the Error/Event Queue as part of this task’s normal processing. The DPU-CCM shall collect a TASK\_HBEAT from DPU-SCUI, DPU-CCM, DPU-DCX, DPU-TMALI, and DPU-DPA. Non-responsive tasks will be reported in DPU\_HK. \\
    \midrule
    3 & The DPU-CCM shall collect a TASK\_HBEAT from DPU-SCUI, DPU-CCM, DPU-DCX, DPU-TMALI, and DPU-DPA . Non-responsive tasks will be reported in DPU\_HK. & Control and Monitoring Every time the CCM Control executes, it calls ccmPerProcess() to handle periodic processing responsibilities. Such responsibilities include analog to digital conversion updates, DPU task monitoring, ICU heartbeat message production, and watchdog strobe. The ccmHealthChk() function, called by ccmPerProcess() verifies the execution of other tasks by monitoring the amount of time that has elapsed since each task last reported. Other tasks report their execution to the CCM Control Task by calling the function, ccmTaskReport(), providing their task index. Each task has an expected execution frequency, and if a task does not execute as expected, an error is reported in DPU housekeeping. If the Command Dispatch Task fails to report for an extended period, the DPU will execute a reboot, since it is impossible to command the DPU if this task is not executing, otherwise it will strobe the watchdog. \\
    \midrule
    4 & The DPU-TMALI shall utilize SCM\_DCI\_SR, along with ERRNO provided by DPU-DCI to decode errors and place them on an error queue for DPU-CCM. & Error Collection and Reporting The ccmErrEnq() function tracks the last error reported and its frequency of occurrence. Once an error code has been reported it becomes the previously reported error code maintained by ccmErrEnq(). A repetition count is then incremented for each subsequent, consecutively reported, identical instance of this previously reported error. If this error code is reported more than once in one high-rate housekeeping reporting period, then a special error, S\_ccm\_ERR\_REPEAT is enqueued with the repetition count for the error encoded in the least significant byte. This mechanism effectively reduces the potential for housekeeping telemetry to become flooded with a single repeated error. \\ 
    \midrule
    5 & The DPU-CCM shall process real-time non-deferred commands within B ms of receipt from the ICU or the SCU. & The Command and Control CSC provides the core command and control functionality for the system. It includes tasks for initializing the system at bootup, scheduling housekeeping data generation, monitoring other tasks, executing periodic tasks, and receiving and dispatching real-time commands. It maintains data structures for system state, commands, errors and events. \\
    \bottomrule
  \end{tabularx}
  \label{tab:investigated_links}
\end{table*}

\newpage
\balance
\bibliographystyle{IEEEtranS}
\bibliography{references/related}

\end{document}